\documentclass[]{aastex631}

\usepackage{amsmath}
\usepackage{graphicx}	% Including figure files
\usepackage{amstext}% Advanced maths commands
\usepackage{comment}
\usepackage{derivative}
\usepackage{mathtools}
\usepackage{anyfontsize}
\usepackage{cuted}

\received{July 2, 2024}
\begin{document}

\title[Adaptive Habitability of Exoplanets]{Adaptive Habitability of Exoplanets:\\
Thriving Under Extreme Environmental Change}

\author{Itay Weintraub}
\affiliation{Technion Israel Institute of Technology, Physics Department, Haifa 32000, Israel}
\email{weintraub@campus.technion.ac.il}

\author{Hagai B. Perets}
\affiliation{Technion Israel Institute of Technology, Physics Department, Haifa 32000, Israel}
\affiliation{Department of Natural Sciences, The Open University of Israel, 1 University Road, PO Box 808, Raanana 4353701, Israel}
\email{hperets@physics.technion.ac.il}
\begin{abstract}
The dynamic nature of life's ability to thrive in diverse and changing planetary environments suggests that habitability and survival depend on the evolutionary path and life adaptation to environmental conditions. Here we explore such "adaptive habitability" through astro-ecological models. 
 We study the interplay between temperature adaptation and environmental fluctuations, particularly those induced by solar activity and orbital dynamics. We present a simplified ecological-evolutionary model to investigate the limits of life's adaptability on a planetary scale. By incorporating complexities such as multiple niches, migration, species interactions, and realistic temperature variations, we demonstrate the potential for adaptive habitability in the face of both gradual and abrupt environmental changes. Through simulations encompassing monotonic, periodic, and secular dynamical evolution-induced temperature profiles, we identify critical thresholds for survival and extinction, highlighting the importance of phenotypic variance and dispersal rates in adapting to varying environmental conditions. These findings underscore the significance of considering temporal variations in assessing exoplanet habitability and expanding the search space for potentially habitable worlds.
\end{abstract}
\section{Introduction} \label{introduction}
Habitability, a central concept in the search for extraterrestrial life, remains a topic of debate \citep{article}. While astronomers often focus on Earth-like conditions, astrobiology suggests a wider range of possibilities \citep{Schulze_Makuch_2018}. Life on Earth has proven capable of thriving in extreme environments \citep{doi:10.1089/ast.2007.0137}, raising questions about the true limits of habitability.
This paper explores the concept of "Adaptive Habitability", recognizing that life can evolve and adapt to changing planetary conditions. We focus on temperature adaptation, as temperature is a key variable influenced by solar activity and orbital dynamics.
Previous work has investigated habitability across various parameters, considering both instantaneous and long-term conditions \citep{Schulze_Makuch_2018}. The availability of energy sources and organic compounds are crucial factors, with solar radiation being a particularly dynamic variable. Life's adaptability is key to surviving these fluctuations.
Our aim is to emphasize the role of time-varying habitability, highlighting its dependence on the rate of environmental change relative to the rate of biological adaptation,
e.g. the rate of genetic mutations facilitating the necessary phenotypic changes required to survive following the change in the environmental conditions. 
We consider planets that are observed to have inhospitable conditions for life but might have been more habitable in a past state. By assessing the evolutionary requirements for survival under changing conditions, we expand the search space for potentially habitable exoplanets. In fact, Earth itself, to some extent, can be considered as such an example, with the conditions existing during the epoch of life formation on Earth being distinctly different from those which later evolved and exist today, e.g. the transition following the great oxidation event on Earth.
This paper presents a simplified ecological-evolutionary model to investigate the limits of life's adaptability, adopting a planetary-level ecological model. Starting with fairly simplistic models, it is possible to study whether a planet may have held a habitable state in the past and assess the evolutionary requirement for a given species to survive the environmental variation.
Such models were explored on small scales and for small changes in small ecological systems on Earth. For example, ecologists have made attempts to study the changes in biological communities under climatic change in the near future. Here we introduce a somewhat similar approach but explore it at the planetary-scale level, and for longer evolutionary times, motivated by the possible dynamical evolution of exoplanetary systems.  
We make use of a planetary scale temperature and evolution model adapted from \citep{Akesson_2021}.
We begin by describing our methodology, presenting the basic assumptions and the initial simpler model we consider, and we then progressively incorporate complexities such as multiple niches, migration, species interactions, and realistic temperature variations. We aim to demonstrate the potential for adaptive habitability and its implications for the search for extraterrestrial life.
\section{Overview}
Prior to a detailed model presentation, we outline the fundamental aspects of evolution captured by our models.
  Consider a species that evolved under specific favorable conditions. If it thrives only within a narrow range of these conditions, it cannot easily migrate to dissimilar environments or survive significant environmental changes.  Thus, species persisting in non-constant environments likely possess adaptation mechanisms, which we categorize here as genetic adaptation and spatial migration.
Genetic adaptation involves changes in physical traits over time to better suit the existing environment. If conditions become less favorable, genetic selection can optimize traits for the new circumstances (e.g., developing fur in colder climates).  Spatial migration enables movement to more favorable locations in response to environmental change (e.g., bird migration or using caves for shelter). Both mechanisms can ensure a species' survival when its environment changes.
A species' survival in a changing environment depends on the relative rates of environmental change, spatial migration (if suitable locations exist), and genetic evolution.  Rapid environmental change may outpace adaptation, leading to extinction.
Life may emerge and evolve under specific conditions (e.g., certain temperatures or oxygen levels), then change over time due to environmental shifts like global warming. The resulting species and its distribution may differ greatly from the original, yet life persists in the new conditions. However, if these "final" conditions had existed initially, life might not have formed at all. Therefore,
a planet's habitability depends not only on its current conditions but also on the history of environmental changes \emph {and} the adaptations of life to these changes.
We term this concept "adaptive habitability".
The rest of this paper details our methodology for exploring these ideas, using simplified models to demonstrate such evolutionary scenarios.
\section{Methods} \label{methods}

\subsection{Basic Ecological Model} \label{Ecological Base Model}

In order to capture the most fundamental behaviors of a community in a single habitat, two state variables are needed - the first is the population density $n$, initially normalized to $n_0=1$, and the optimal trait $m$, in this case, the temperature, which is initially set to the environmental temperature prior to climate change onset. 
For each state equation, several parameters must be defined.
First is the width of the species response function to temperature $T$, which is also referred to as the temperature tolerance $\sigma$, given in units of $(^\circ C)$.
The population growth rate should follow a Gaussian distribution around the optimal trait $m(t)$ with variance $\sigma^2$:
\begin{gather}
        f(t) = \rho \cdot \frac{e^{-(T(t)-m(t))^2 / (2\cdot \sigma^2)} }{\sigma}\label{eq:basic_f}
\end{gather}
Such that reproduction will be encouraged as the species populates a near-optimal habitat. $\rho$ is a factor given in units of $C^\circ yr^{-1}$ which controls the maximal growth rate. It is expected that for a fixed value of $\rho$, a large $\sigma$ will allow the species to survive a wider range of conditions, at the expense of growth rate. The introduction of the growth term $f(t)$ is sufficient to define the differential equation for $n$:
\begin{gather}
     \frac{dn}{dt} (t) = n(t) \cdot [f(t)-\kappa] \label{eq:basic_n}
\end{gather}
With $\kappa$ being an intrinsic mortality rate. In case $|T-m|\gg\sigma$ the growth term is diminished, and natural mortality will eventually lead to extinction. For computational purposes, a minimal population density bar must be set $n_{min}$ for a species to be considered extinct.
For the state equation of the optimal trait $m$, a dependence on the population density must be set since genetic variety is correlated with the size of the community \citep{Hague2016}. A certain genetic change amplitude $v$ should be defined. Additionally, the trait lag, defined as $T_{lag} = T_{env} - m$, should be normalized by the temperature tolerance $\sigma$.  
When $\sigma \gg T_{lag}$ the species is able to tolerate the change without significant adaptation. When climate change exceeds several $\sigma$'s, then the species' survival will depend on $v$, the capability to respond, genetically, to swift fast climate changes.
\begin{gather}
         \frac{dm}{dt} (t) = f(t) \cdot v \cdot \frac{T(t)-m(t)}{\sigma^2}\coloneqq g(t)\label{eq:basic_m} 
\end{gather}
In order to understand why the growth rate $f(t)$ is participating in equation \ref{eq:basic_m} instead of $n(t)$, it is preferable to trace back the original formalism for this equation from \cite{Norberg_2012} (Which itself is applying the model of \cite{Kirkpatrick1997-lw})
\begin{gather}
    \frac{dm}{dt} (t) = v \cdot \pdv{\Tilde f(t)}{m} \label{eq:m_withFitness}\\
    \Tilde f(t)= f(t)-\kappa \label{eq:fitness},
    \end{gather}
where $\Tilde f(t)$ is the effective growth rate of $n$, also termed the "fitness function" in the ecological context. The right-hand side of equation \ref{eq:m_withFitness} pronounces the optimization with respect to m, over a normal distribution. This phenomenon is well-known in ecology and is termed "directional selection".
\subsection{Full Ecological Model}\label{Full Ecological Model}
Besides the basic model described above we also consider a more complex and more realistic model which we describe below. In particular, several simplifications made by the basic model were relaxed.
First, spatial dependency was included. The planet was divided into $L=20$ habitats, and so $n$, $m$, and $T$ depend on the spatial position, $x$. As different environmental conditions are present in the model, the species are capable of migrating from one habitat to its neighbors. This allows for a certain level of non-genetic adaptation response of the species to the changing climate, through migration to different spatial niches on the planet.  
The second major change with respect to the basic model is allowing for multiple species and the competition between them. For simplicity and computational speed, only $s=4$ species were simulated.
Both migration and competition are then included in the state equation for both $n$ and $m$.
Migration is characterized by a mean (over species) spatial dispersal rate $d$. The genetic variance $v$ is now slightly varying between species as well, in order to consider a (weak) selection work frame.
It should be disclaimed that the term 'species' considered here serves as a simplified parameterized entity rather than a detailed biological definition. The term 'species' refers to all communities that share the same properties, namely $v$, $d$ and $\rho$. For each spatial niche, a species is allowed to have a different $m$ since $T(x, t)$ now depends on the position on the planet, and is generally non-homogeneous over the entire planet. 
The fair reasoning is to think of the simulated species as a phylogenetic tree of its own, originating from a common ancestor. As species adapt and migrate, their biological identity may vary, while their dispersal and phenotypic variances are kept.
The model described here is based to some extent on the complex ecologic model described in \citep{Akesson_2021}, which we briefly 
define and describe here, with the appropriate changes and differences relevant to our exploration; we refer the reader interested in a more detailed explanation to \citep{Akesson_2021}
The equations will be arranged in a top-down manner, starting at the new state equation for the population density $n$ and the optimal temperature trait $m$:
\begin{gather}
    \frac{dn^i_k}{dt} (t) = S(\frac{n^i_k (t)}{n_{min}})\cdot \underbrace{[n^i_k (t)  [f^i_k (t)-\kappa]}_\text{Intrinsic Growth}-\underbrace{\gamma^i_k (t)}_\text{Competition}] + \underbrace{\mu^i_k (t)}_\text{Migration}   \label{eq:dndt}\\
    \frac{dm^i_k}{dt} (t) = h^i_k (t)^2 \cdot [ \underbrace{g^i_k (t)}_\text{Directional Selection}- \underbrace{\Tilde{\gamma}^i_k (t)}_\text{Competition} + \underbrace{\Tilde{\mu}^i_k (t)}_\text{Migration}] \label{eq:dmdt}
\end{gather}
For all equations on this model, indices $i$ and $j$  run over species, and $k$ runs over spatial niches. Both state variable equations are composed of a  generalization of the base model, in addition to a migration term and a competition  term.
Diving in the population density state equation, the intrinsic growth is given by:
\begin{gather}
    f^i_k (t)= \rho^i \cdot \frac{e^{-(T(\frac{k}{L},t)-m^i_k (t))^2 / (2\cdot (\sigma^i_k (t))^2)}}{\sigma^i_k (t)} \label{eq:f} \\
    \sigma^i_k (t)= |b-a\cdot m^i_k (t)|\label{eq:sigma} 
    \end{gather}
The non trivial addition compared to the basic model is the temperature tolerance $\sigma$  dependence on the optimal trait $m$. 
  $a,b$  control the trade-off between maximal growth rate and temperature tolerance. For colder temperatures the tolerance is slightly higher, on the expense of diminished maximal growth rate, while for warmer habitats the growth rate is amplified in exchange for smaller tolerance (see \citep{Akesson_2021} for  elaboration).
Addressing the new terms in the population density equation, one should expect the migration term to follow a simple diffusion equation form, and this is indeed the case, in discrete manner:
\begin{gather}
        \mu^i_k (t)= d^i \cdot \sum_l \underbrace{M_{kl} \cdot n^i_l (t)}_\text{immigration} - \underbrace{M_{lk} \cdot n^i_k (t)}_\text{emigration} \label{eq:mu}
        \end{gather}
$M$ is a migration matrix, 1 for neighboring niches ($l=k\pm1$) and zero otherwise. Re-organizing the terms will yield the familiar discrete second derivative with respect to location.
The population loss due to inter-species competition on resources is described through a competition matrix $\alpha^{ij}_k$:
\begin{gather}
        \gamma^i_k (t)= \sum_j n^i_k (t) \cdot \alpha^{ij}_k \cdot n^j_k (t) \label{eq:gamma}\\
        \alpha^{ij}_k = e^{-\Delta ({m^{ij}_k})^2 /\eta^2}\label{eq:alpha} \\
    \Delta m^{ij}_k := m^j_k - m^i_k\label{eq:dm}
\end{gather}
Equation \ref{eq:alpha} describes a trait difference dependent competition. That is, the closer two species are in the basic description of their optimal habitat, which in our case characterized by temperature, more significant competition will take place, and so larger population penalties will follow. The severity of competition is also governed by the competition width $\eta$. A larger competition width means species with a wider range of trait values compete strongly, while a smaller competition width means only species with very similar traits experience significant competition. 
Lastly, right next to the generalized growth term we first encounter $S$, which is the approximation to smooth step function described in equation ~\ref{eq:smoothstep}.
$n_{min}$ is a threshold value for population density below which the change rate is diminished.
Now that multiple habitats and species are allowed, extinction occurs only where $n<n_{min}$ for all species on the entire planet.
Therefore, partial extinction does not stop the model from following the dynamics of species that did not fail. The smooth step function is used to avoid discontinuity when using an ODE solver, but it effectively ignores the extinct communities.  Notice that $S$ factors all terms except migration, since immigration to the habitat is allowed after the local community became extinct.
Now that the population density equation is fully addressed, drawing the equivalent terms for the optimal trait equation should be more manageable.
First, the generalization of the directional selection described for the base model in equations \ref{eq:basic_m}-\ref{eq:fitness}:
\begin{gather} 
    g^i_k (t) = \frac{f^i_k (t) \cdot v^i \cdot (T(\frac{k}{L},t) -m^i_k (t))}{(\sigma^i_k (t))^2}\label{eq:g}
\end{gather}
The competition term in the current context represents the genetic update each species exhibits in reaction to rival species on a shared habitat:
\begin{gather}
          \Tilde{\gamma}^i_k (t)= \sum_j \beta^{ij}_k \cdot n^j_k (t) \label{eq:tildgamma}\\
          \beta^{ij}_k = v^i \cdot \pdv{ \alpha^{ij}_k}{(\Delta m^{ij}_k)}=\frac{2v^i \cdot \Delta m^{ij}_k}{\eta^2} \cdot\alpha^{ij}_k \label{eq:beta} 
\end{gather}
Since competition is also viewed as a selection force in ecology, the 
trait update related to competition is a derivative of the equivalent for population update with respect to $\Delta m$, multiplied by the genetic variance $v$, same as the relation between $f$ and $g$. 
As for the migrative gene flow, the resulting term is again a discrete second derivative of a diffusion equation, in resemblence to equation \ref{eq:mu}:
\begin{gather}
     \Tilde{\mu}^i_k (t)= d^i \cdot \frac{\sum_l M_{kl} \cdot n^i_i (t) \cdot (m^i_l (t) - m^i_k (t) )}{n^i_k (t) +\epsilon} \label{eq:tildmu}
\end{gather}
$\epsilon$ is a floor density value in order to avoid infinite $\Tilde{\mu}$.
Lastly, the equivalent to the step function term $S$ is now the within the heritability. In ecology, heritability is the ratio between the selection differential and the response to selection, according to the breeder's equation, and is marked by $h^2$. For the single species model, the heritability was a constant, set to 1 for simplicity. In this model:
\begin{gather}
    h^i_k (t)^2 = \frac{q^i_k (t) }{q^i_k (t) +v}\label{eq:h}
\end{gather}
Where:
\begin{equation}
        q^i_k (t) = v^i \cdot S(n^i_k (t)/n_{min}),\label{eq:q} 
\end{equation}
with the use of the already familiar $S$ function.
Altogether, one can notice the two state equations for $m,n$ are strongly correlated, and both are a direct extension of the base model. The population density equation is composed of a temperature-lag dependent intrinsic growth term, competition related losses, and migration from neighboring habitats.
The optimal trait equation is composed of a temperature-lag dependent directional selection,  competition selection and migrative genetic mixing.
\subsection{Temperature Profiles} \label{Temperature Profiles}
Let us first define the general form of a temperature profile in our model:
\begin{gather}
            T(x,t)=(T_{max}-T_{min})\cdot x+T_{min}+((C_{min}-C_{max})\cdot x + C_{max})\cdot T(\frac{t}{t_E}) \notag
	\label{eq:temprature_xt}
\end{gather}
$T(x,t)$ is the temperature for each niche and time step of the simulation.
$T_{max}$ and $T_{min}$ are the equatorial and polar initial temperatures, respectively.
$x$ is the discrete normalized latitude over half a sphere, when $x=0$ for the pole, and $x=1$ for the equator. 
$C_{max}$ and $C_{min}$ are the polar and equatorial total temperature changes, respectively. 
$t_E$ is the full duration of the climatic change. The simulation starts at a certain adaptation phase, such that $t=0$ is the climatic change starting time.
The temperature temporal profile can be defined either as a smooth-step approximation:
\begin{equation}
    T_{smooth}(t) = \begin{cases}
        0   &t\leq0 \\ 
        6t^5 -15t^4 + 10t^3 &0<t<1  \\
        1   &t=1
        \label{eq:smoothstep}
    \end{cases}
\end{equation}
Or a sinusoidal function;
\begin{equation}
    T_{periodic}(t) = \begin{cases}
        sin(2P \pi t)  &0\leq t \leq1  
        \label{eq:sin}
    \end{cases}
\end{equation}
For a certain period number $P$.
\subsection{von-Ziepel-Kozai-Lidov Oscillations Temperature Profile} \label{Kozai-Lidov Oscillations Temperature Profile}
In addition to the ideal climate change profiles described in equations \ref{eq:smoothstep}  and \ref{eq:sin}, a more astrophysical-rooted temperature constraint was derived from a simulated von-Ziepel-Kozai-Lidov (vZLK) secularly evolving planetary system [\citep{https://doi.org/10.1002/asna.19091832202}, \citep{LIDOV1962719} ,\citep{1962AJ.....67..591K}].
vZKL evolution is a secular process occurring in hierarchical three-body systems, mostly pronounced in systems in which a large mutual inclination exists between the outer orbit of the third body with respect to the inner binary orbit.
On timescales much larger than the orbital periods, the eccentricity and inclination of the inner bodies exhibit periodic changes due to angular momentum exchange between the two orbits.
The full Hamiltonian of the problem can be described as a sum of the gravitation Hamiltonian of each orbit, plus an interaction term.
The interaction energy between the two orbits can be written as a series expansion. Each term is a function of the three bodies masses ($m_1, m_2,m_3$), the major axis of the orbits ($a_1,a_2$), the angle between the two angular momenta ($\Phi$), and the distance of the bodies from system's center of mass ($r_{in},r_{out}$).
Taking a finite number of terms within the expansion yields reasonable approximations as well. The first contributing order is the quadruple, with the octupole order next in line.
In order to solve the Hamiltonian several other approximations are made. First, the secular approximation, meaning averaging over the orbital period (smallest period timescale), nullifies the dependence on specific body distance in respect to $a_1$ or $a_2$, which in term stays constant.
Another simplifying step is to take $m_1$ or $m_2$ to zero, also known as the test particle approximation. Lastly - the axisymmetric potential approximation is taking the orbits to be nearly circular ($e_1, e_2 \rightarrow 0$).
The vZLK model used follows the description in the supplementary materials of \citep{Naoz_2016}. The equation set is a full octupole order treatment, relaxing the test particle and axisymmetric potential approximations.
The tested 3-body hierarchical system followed a Sun-Earth-Jupiter scheme, with proper modifications in order to yield a non-chaotic planetary eccentricity from one hand, but significant enough eccentricity changes so the biological community will be challenged by the temperature shift.
In our context, the inner binary is the star (object 1) and Earth-like planet (object 2), while the outer perturber is a heavy Jupiter-like planet (object 3).
The degrees of freedom of the vZLK simulation are
the masses of the three objects ($m_{1,2,3}$) and the major axis for inner and outer orbits ($a_{1,2}$ ) which are constant under the secular averaging of the planetary dynamic.
The variables that need to be set at the start of the vZLK simulation are:
$i_{1,2}$ inclinations of inner and outer orbits,
$\omega_{1,2}$ argument of periapsis for inner and outer orbits, 
$G_{1,2}$ angular momenta of inner and outer orbits and
$e_{1,2}$ eccentricities of inner and outer orbits
The only vZLK output relevant for our purposes is the eccentricity of the Earth-like orbit, $e_1$. The time evolution of $e_1$, together with the given obliquity of the planet, is enough to derive the temperature over time in two points on the planet - the poles and the equator.
The relation between these parameters and the temperature is given as follows:
\begin{gather}
     U = L_s \cdot \frac{1-a}{16\cdot \sigma \cdot \pi d^2} \cdot (1-e^2)^{-0.5}\label{eq:U}\\ 
    U_{p} = \frac{1}{\pi} \cdot U \cdot sin(\epsilon)\label{eq:Up} \\
    U_{eq} = \frac{2}{\pi ^2} \cdot U \cdot \Tilde{E}(sin(\epsilon))\label{eq:Ueq} \\ 
    \Tilde{E}(x) = \int_{0}^{\frac{\pi}{2}} \sqrt{1-x^2 sin^2\theta} d\theta \label{eq:ellipticalInt} \\
    T_{p, eq} = \sqrt[4]{U_{p, eq}} \label{eq:TfromU}
\end{gather}
Where $L_s$ is the star's luminosity, $a$ is the surface albedo of the planet, and $\sigma$ is the Stephan-Boltzmann constant.$e$ is the planet's eccentricity.
$d$ is the average star-planet distance, taken as a first approximation to be the major axis of the planet's orbit.
$U$ is the annually averaged solar irradiation per area absorbed by the planet's surface.
$U_{p, eq}$  are the portions of $U$ absorbed on the polar and equatorial regions alone, respectively. 
$\epsilon$ is the planetary obliquity. $\Tilde{E}(x)$  is $E(\frac{\pi}{2},x) $, the complete elliptical integral of the second type, as described by equation \ref{eq:ellipticalInt}.
$T_{p, eq}$ are the derived temperature constraints on the planet, which are later used to linearly interpolate temperature between the poles and the equator for simplicity.
The relation between obliquity, eccentricity and the annually averaged solar irradiation can be found in \citep{PAILLARD2010273}. 
\section{Results} \label{results}
As a first step, we demonstrate the behavior of the basic model and the relations between the key variables in the system described in subsection \ref{Ecological Base Model}.
The climatic variation was chosen to $\Delta T = 10 C^{\circ}$ over a time-span of $\Delta t = 1\cdot10^6$ years, on a rising step manner as indicated by equation \ref{eq:smoothstep}.
The remaining parameters used are listed in table \ref{tab:ecol_base_params}.

\begin{table}[ht]
\centering
\caption{Ecological Base Model Parameters}
\label{tab:ecol_base_params}
\begin{tabular}{|l|l|l|}
\hline
\textbf{Parameter} & \textbf{Symbol} & \textbf{Value} \\
\hline
Temperature variation amplitude & $\Delta T$ & $10 \, ^\circ\text{C}$ \\
Temperature variation timescale & $\Delta t$ & $1 \times 10^6$ years \\
Temperature tolerance& $\sigma$ & $2 \, ^\circ\text{C}$ \\
Intrinsic mortaility rate& $\kappa$ & $1 \times 10^{-4}$ yr$^{-1}$ \\
Growth-tolerance trade-off coefficient & $\rho$ & $2 \times 10^{-4}$ yr$^{-1}$ \\
Initial temperature& $T_0$ & $15 \, ^\circ\text{C}$ \\
Minimal viable population density & $n_\text{min}$ & $1 \times 10^{-5}$ \\
\hline
\end{tabular}
\end{table}

\begin{figure}[ht!]
    \centering
    \includegraphics[width=1\linewidth]{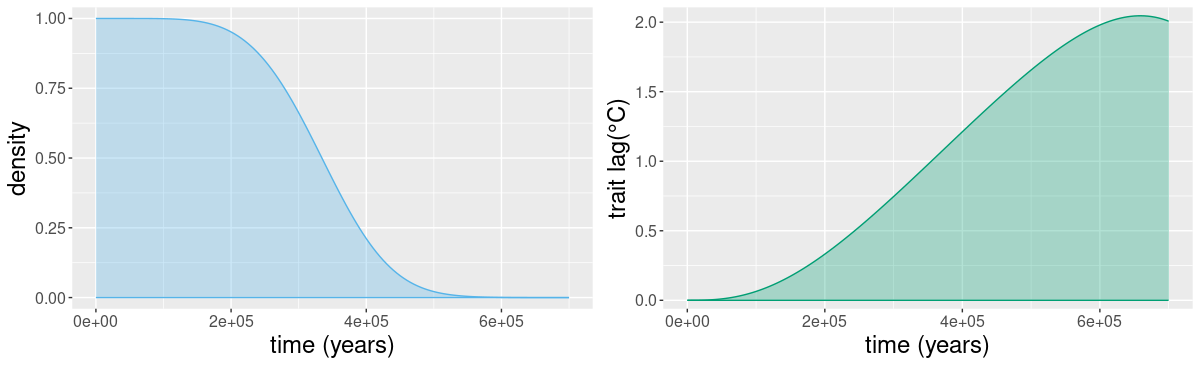}
    \includegraphics[width=1\linewidth]{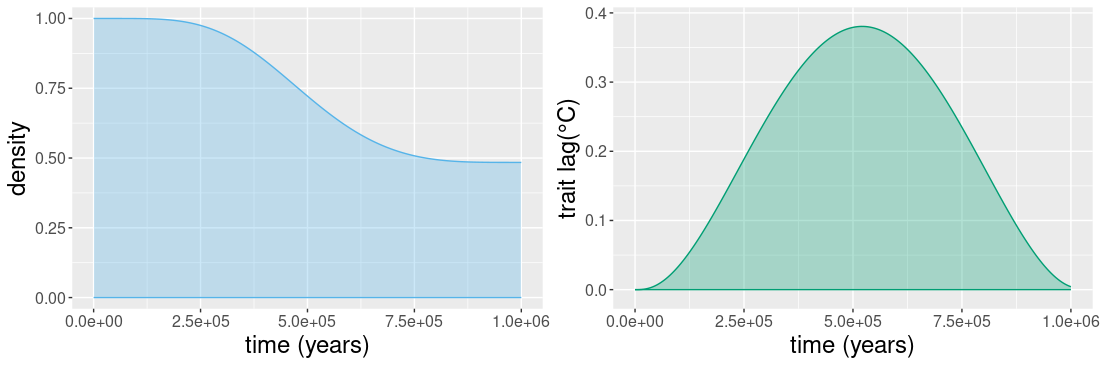}
    \caption{ Population density and trait lag over time for a rising step climatic change of $10 C^{\circ}$. The lower panel has 4 times the genetic change amplitude $v$ compared to the upper panel.}
    \label{fig:firstDemo}
\end{figure}

Figure \ref{fig:firstDemo} shows the population density $n$ and the trait lag $T - m$ over time in a single habitat, under the pre-mentioned climatic change. For the upper panel, the ratio between  the genetic change amplitude $v$ and squared value of the temperature tolerance $\sigma^2$  is 0.125, while for the lower panel, the ratio is taken to be equal to 0.5.

Two interesting extreme cases of the basic model are ideal adaptation and no adaptation. Those cases translate to $v\rightarrow \inf$ and $v\rightarrow 0$, respectively.

For the ideal adaptation case, the parameter set used is listed in table \ref{tab:ecol_base_params_ideal}.

\begin{table}[ht]
\centering
\caption{Ecological Base Model Parameters (Ideal Adaptation)}
\label{tab:ecol_base_params_ideal}
\begin{tabular}{|l|l|l|}
\hline
\textbf{Parameter} & \textbf{Symbol} & \textbf{Value} \\
\hline

\textbf{Ideal Adaptation Parameters} & & \\
\hline
Variance of phenotypic trait& $v$ & $1 \times 10^{10} \, (^oC)^2$ \\
Temperature tolerance& $\sigma$& $1 \times 10^{-2} \, ^oC$\\
Growth-tolerance trade-off coefficient& $\rho$ & $\kappa \cdot \sigma \cdot r$ yr$^{-1}$ \\
\hline
\end{tabular}
\end{table}

\begin{figure}[ht!]
    \centering
    \includegraphics[width=1\linewidth]{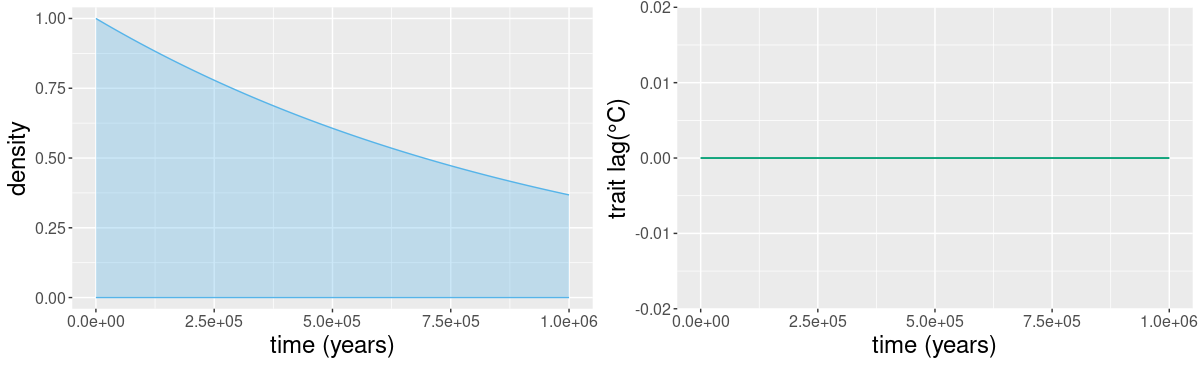}
    \includegraphics[width=1\linewidth]{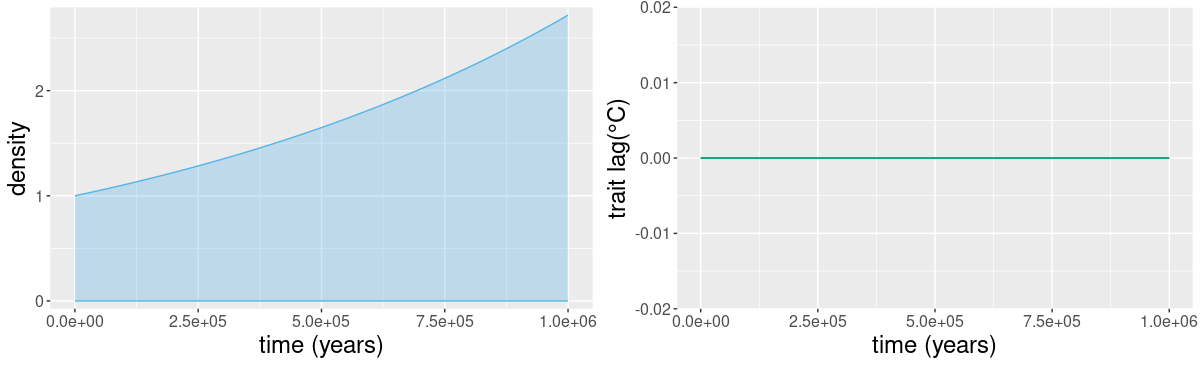}
    \caption{Population density and trait lag over time for ideal adaptation edge case $v\rightarrow \inf$, $\sigma \ll C$. The lower panel has $2\%$ bigger maximal growth rate $\rho$, compared to the upper 
    panel.}
    \label{fig:secondDemo}
\end{figure}
Where $r$ is a scaling factor. This set of parameters forces immediate follow-up on the environmental condition, without relying on any temperature tolerance such that $\sigma<< C$, With the other parameters unchanged. Figure \ref{fig:secondDemo} shows the population density and trait lag over time. The left panel shows $r=1.01$, and the right panel shows $r=0.99$. It is clear that given ideal adaptation, there is an exponential sensitivity around $r=1$.

The opposite edge case is $v = 0$, and $\sigma \gg C$, which stands for total dependence on the species temperature tolerance, without any directional selection ($m$ is forced to be constant). In this case, $r$ was set to be 1.01 so a certain initial growth will be noticeable.

\begin{figure}[ht!]
    \centering
    \includegraphics[width=1\linewidth]{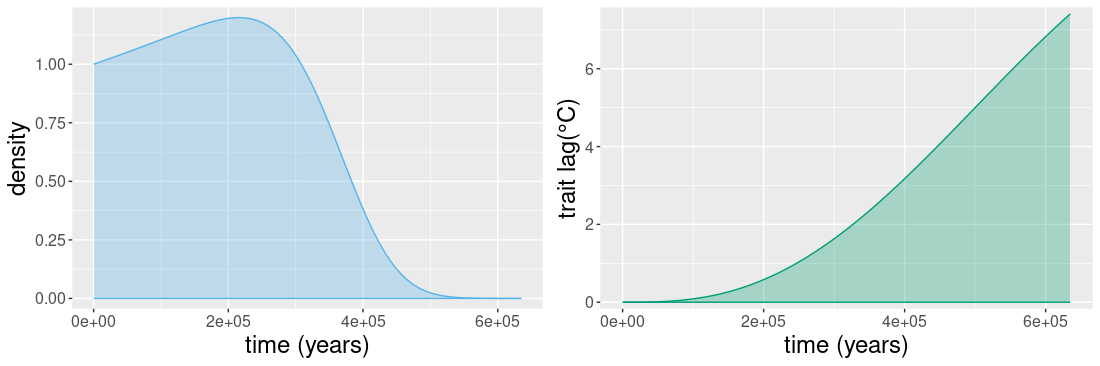}
    \includegraphics[width=1\linewidth]{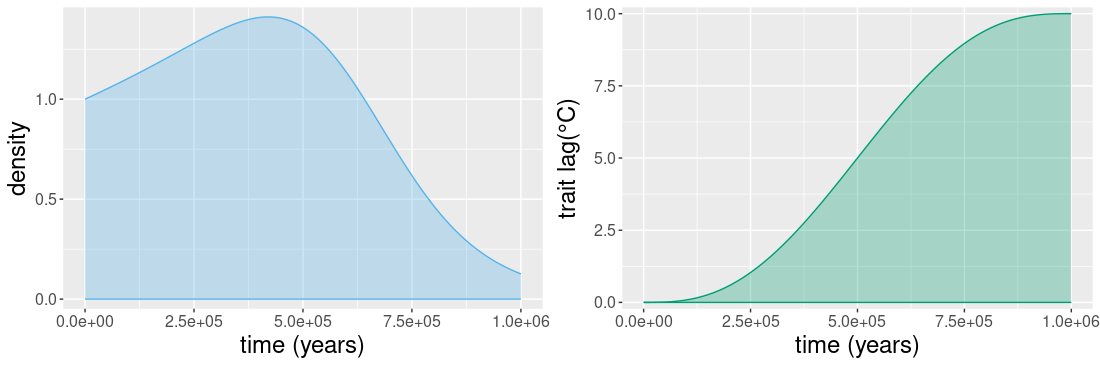}
    \caption{ Population density and trait lag over time for no adaptation edge case $v=0$. Lower panel has $\sigma = 25 > C$ while upper panel has $\sigma = 5 < C$.}
    \label{fig:ThirdDemo}
\end{figure}

On the upper panel of figure \ref{fig:ThirdDemo} , $\sigma = 5 < C$ .On the lower panel $\sigma = 25 > C$. It appears that sufficient temperature tolerance allows a species to survive the entire climatic change, even without any adaptation at all.

After demonstrating the behavior of the simplified model, we test the full model, by varying a small subset of the available parameters - pre-climatic change adaptation time, and nominal phenotypic variance $v$. Since the role of $v$ on the base model was explained, it is more proper from this point on to use the term for $v$ used by \citep{Akesson_2021}.
As indicated in section \ref{methods}, three temperature profile configurations were tested. For each, a survival scenario and an extinction scenario were found.

\subsection{Monotonically rising temperature}
\begin{figure}[ht!]
    \centering
    \includegraphics[width=1\linewidth]{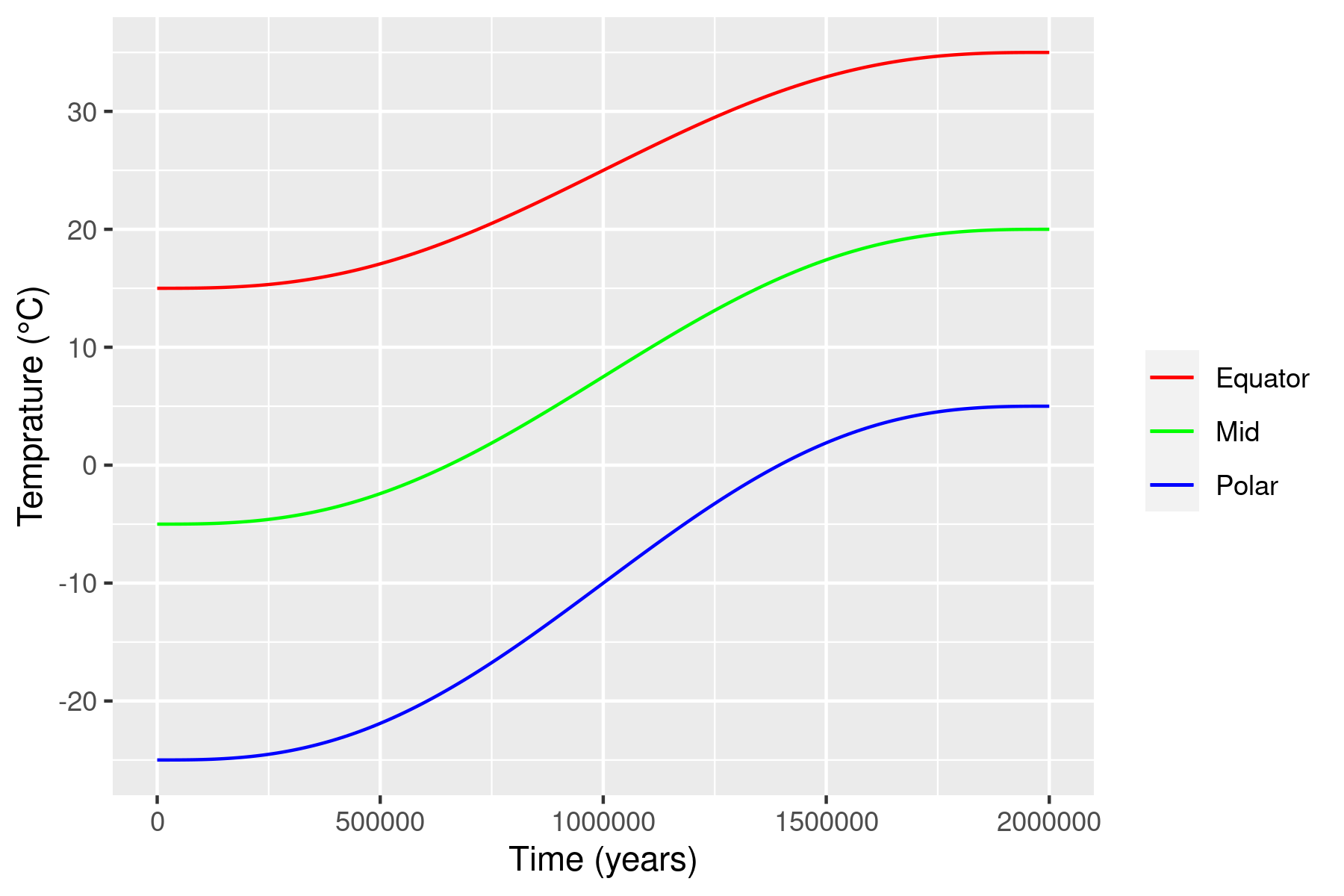}
    \caption{Step Function Temperature Profile}
    \label{fig:stepTemp}
\end{figure}

The simplest case is a rising step function, with the pre-climate change preparation time as the main degree of freedom. 
The profile is drawn in figure \ref{fig:stepTemp} for three loci: Polar, Equatorial, and 45$^\circ$ Parallel environments (symmetrical description for north and south planetary hemispheres).
As described in equation \ref{eq:temprature_xt}, for each timestamp the climate is linearly varying between the poles and the equator, but the total temperature rise is uneven throughout time, with larger changes at the poles.

In the following cases, a nominal phenotypic variance of $v=6\cdot 10^{-5} (C^{\circ})^2$  and a nominal dispersal rate of $d = 1\cdot 10^{-7} (\frac{\pi}{2} \frac{ R_{planet}}{year})$ were chosen, in order to match the climate change timescale properly. The full ecological parameter set ,with variation over species,  is listed in table \ref{tab:ecol_params_full}.

\begin{table}[ht]
\centering
\caption{Ecological Full Model Parameters }
\label{tab:ecol_params_full}
\begin{tabular}{|l|l|l|}
\hline
\textbf{Parameter} & \textbf{Symbol} & \textbf{Value} \\
\hline
 \textbf{Global parameters}& &\\
 \hline
 Temperature tolerance parameters& a&$1 \times 10^{-1}$ \\
 & b&$4\, ^oC$\\
 Competition width& $\eta$&$1\, ^oC$\\
 Intrinsic mortality rate& $\kappa$&$1 \times 10^{-1}\, yr^{-1}$\\
 & &\\
\hline

\textbf{Monotonically rising temprature parameters}& & \\
\hline
Variances of phenotypic trait& $v_{1,2,3,4}$& $6.59, 6.3,6.46,6.46\times 10^{-5} \, (^oC)^2$\\
Dispersal rates& $d_{1,2,3,4}$& $1.03,1.01,1.07,1.12 \times 10^{-7} \, (\frac{\pi}{2} \frac{ R_{planet}}{year})$\\
Growth-tolerance trade-off coefficients& $\rho_{1,2,3,4}$& $6.41, 9.05,4.05,2.17\, yr^{-1}$\\
\hline
 \textbf{Periodic climate change parameters}& &\\
 \hline
 Variances of phenotypic trait & $v_{1,2,3,4}$&$6.35, 7.12, 7.38,6.12\times 10^{-5} \, (^oC)^2$\\
 Dispersal rates & $d_{1,2,3,4}$&$1.49,1.97, 1.52,1.98 \times 10^{-7} \, (\frac{\pi}{2} \frac{ R_{planet}}{year})$\\
 Growth-tolerance trade-off coefficients& $\rho_{1,2,3,4}$&$6.85, 10.11,1.09,5.2\, yr^{-1}$\\
 \hline
 \textbf{Secular dynamical evolution induced climate change parameters}& &\\
 \hline
 Variance of phenotypic trait - Low& $v$&$3.23 \times 10^{-5} \, (^oC)^2$\\
 Variance of phenotypic trait - High& &$6.45 \times 10^{-5} \, (^oC)^2$\\
 Dispersal rate & $d$&$1 \times 10^{-3} \, (\frac{\pi}{2} \frac{ R_{planet}}{year})$\\
 Growth-tolerance trade-off coefficients& $\rho_{1,2,3,4}$&$2.83, 10.2,5.65,8.81\, yr^{-1}$\\
\hline
 \end{tabular}
\end{table}

The intuitive description for the dispersal rate calibration is $d=1$ for covering the distance from the pole to the equator in 1 year.  For an Earth-sized planet, the chosen dispersal rate is approximately $1 \frac{m}{year}$.

\begin{figure}[ht!]
    \centering
    \includegraphics[width=1\linewidth]{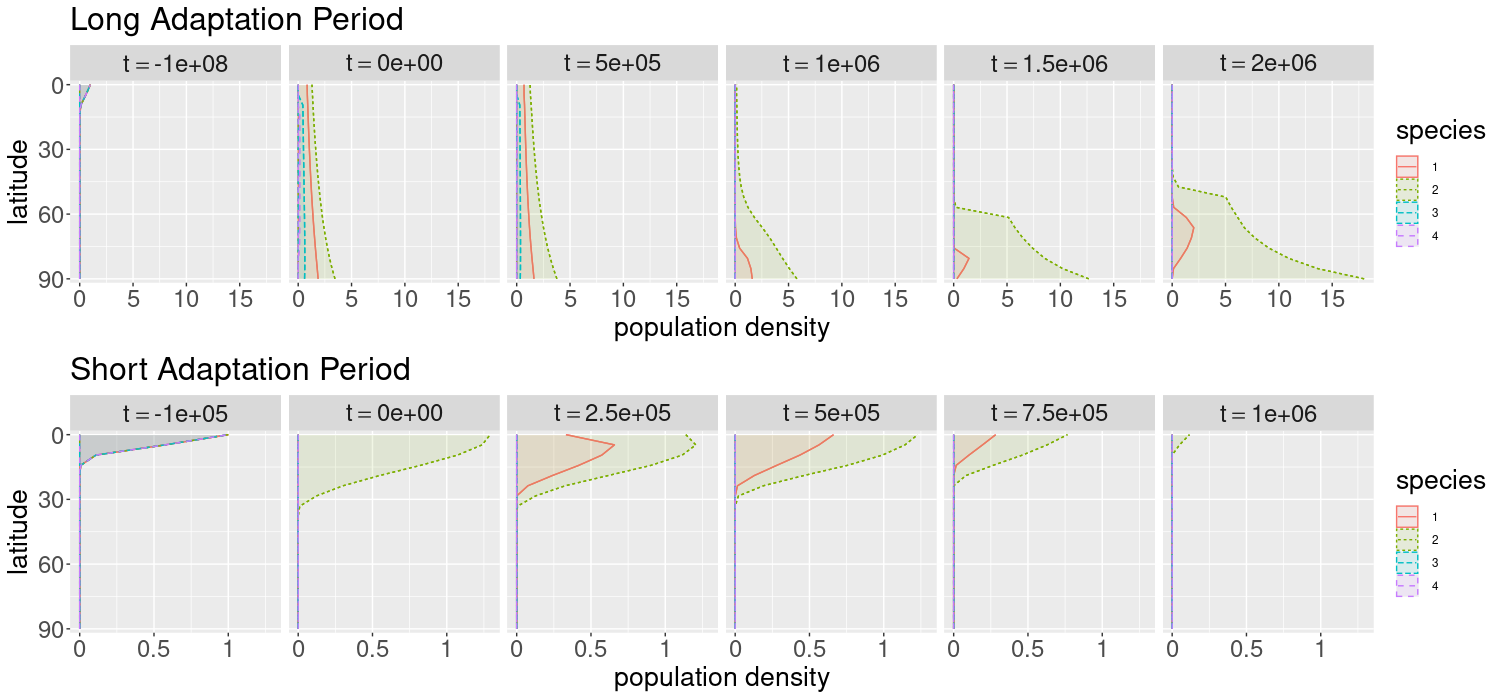}
    \caption{1-D spatial distribution of population density in a rising step climate change, for long and short adaptation times.}
    \label{fig:stepDensity}
\end{figure}

All participating species were generated at the pole and were given an optimal trait $m$ matching their starting environment. We considered two extreme cases, one allowing for species to adapt during a significant time before introducing temperature changes, and one in which temperature changes were introduced shortly after the initialization. 
In the first case (which resulted in species survival at the end of the simulation), the model was evolved for a preparation time of $1\cdot10^{8}$ years, in which no temperature change was introduced, such that the species had sufficient time  to migrate and adapt to other environments, before a temperature change was introduced. 
In the second case (which eventually resulted in full extinction of the species), only a short preparation time of $1\cdot10^{5}$ years was given, not allowing for sufficient time for migration and adaptation throughout the planet.
The population density evolution throughout time is depicted in figure \ref{fig:stepDensity} for both the long and the short adaptation times.

The population density is displayed across the 1-D planetary representation for the chosen time points.
At the start of the simulation, four species were defined with different traits (different relative levels all at an order of 1) with respect to the nominal dispersal rate and nominal phenotypic variance.
Additionally, the species differ in their efficiency of utilizing local resources for growth ($\rho$), thus, dominant species could overtake others (as observed early on). See table \ref{tab:ecol_params_full} for the specific parameter values.

\subsection{Periodic climate change}
\begin{figure}[ht!]
    \centering
    \includegraphics[width=1\linewidth]{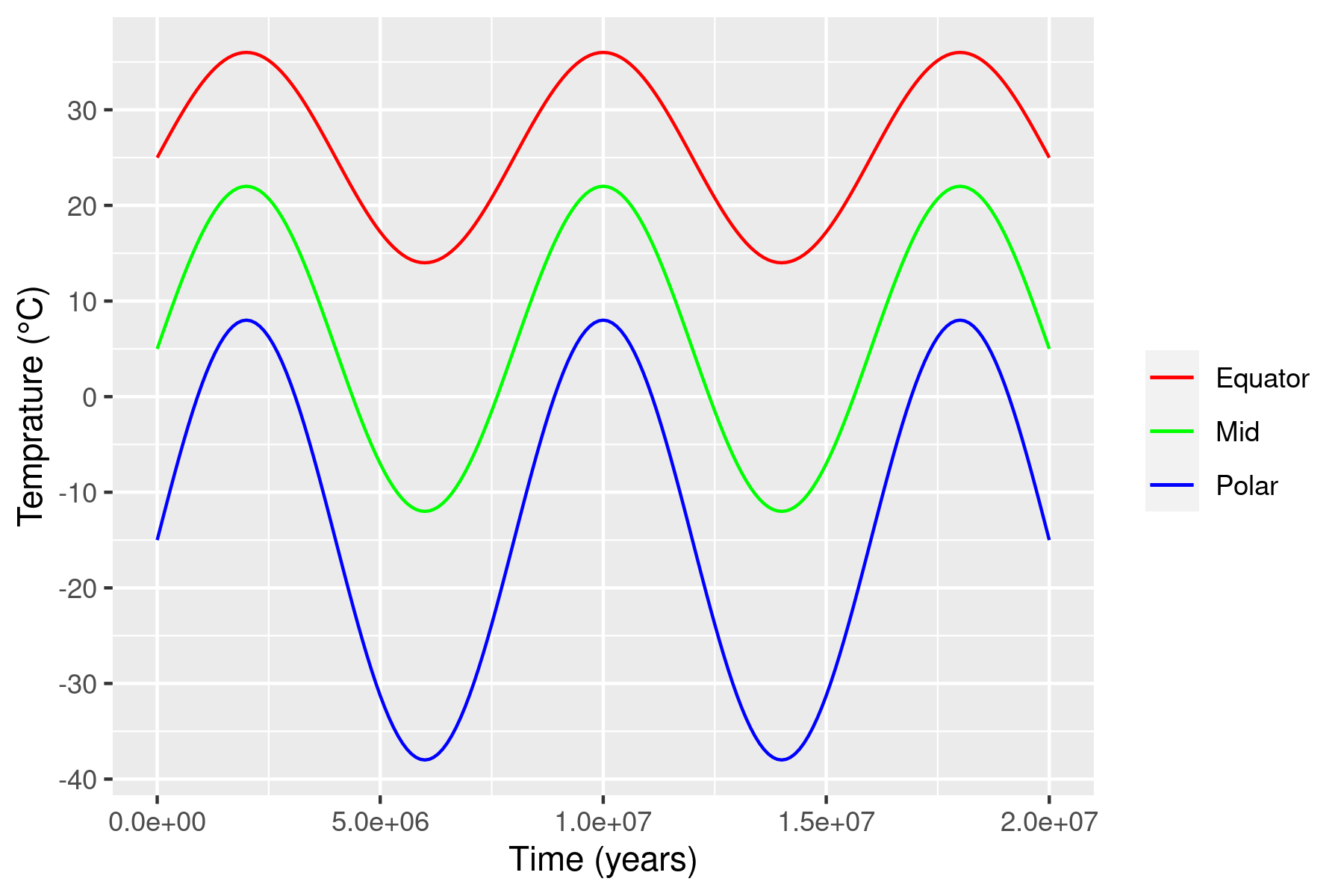}
    \caption{Sinusoidal Temperature Profile}
    \label{fig:sinTemp}
\end{figure}

\begin{figure}[ht!]
    \centering
    \includegraphics[width=1\linewidth]{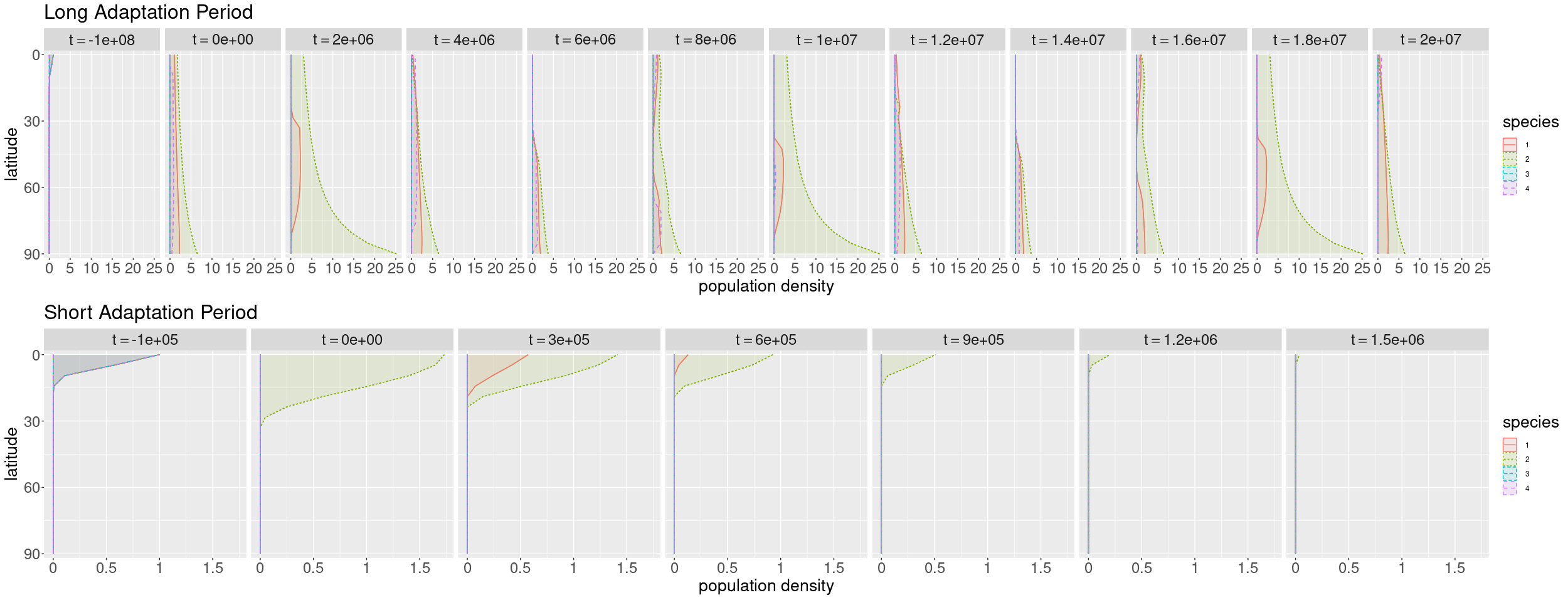}
    \caption{1-D spatial distribution of population density in a periodic climate change, for long and short adaptation times.}
    \label{fig:sinDensity}
\end{figure}

The next profile configuration to be tested was the sinusoidal climate change, depicted in figure \ref{fig:sinTemp} and described by equations \ref{eq:temprature_xt} and \ref{eq:sin}. Similarly to the profile illustrated in figure \ref{fig:stepTemp}, the climatic change is uneven across the planet, with growing amplitude from the equator to the poles.
Bearing the maximal derivative finding in mind, one will concur that for the short adaptation case, the species simulation should fail at the first inflection point, rather than a certain extremum. In contrast, in the long adaptation case, species suffer significant periodic losses but eventually prevail through periodic changes.
The corresponding results are shown in figure \ref{fig:sinDensity}.

\begin{figure}[ht!]
    \centering
    \includegraphics[width=1\linewidth]{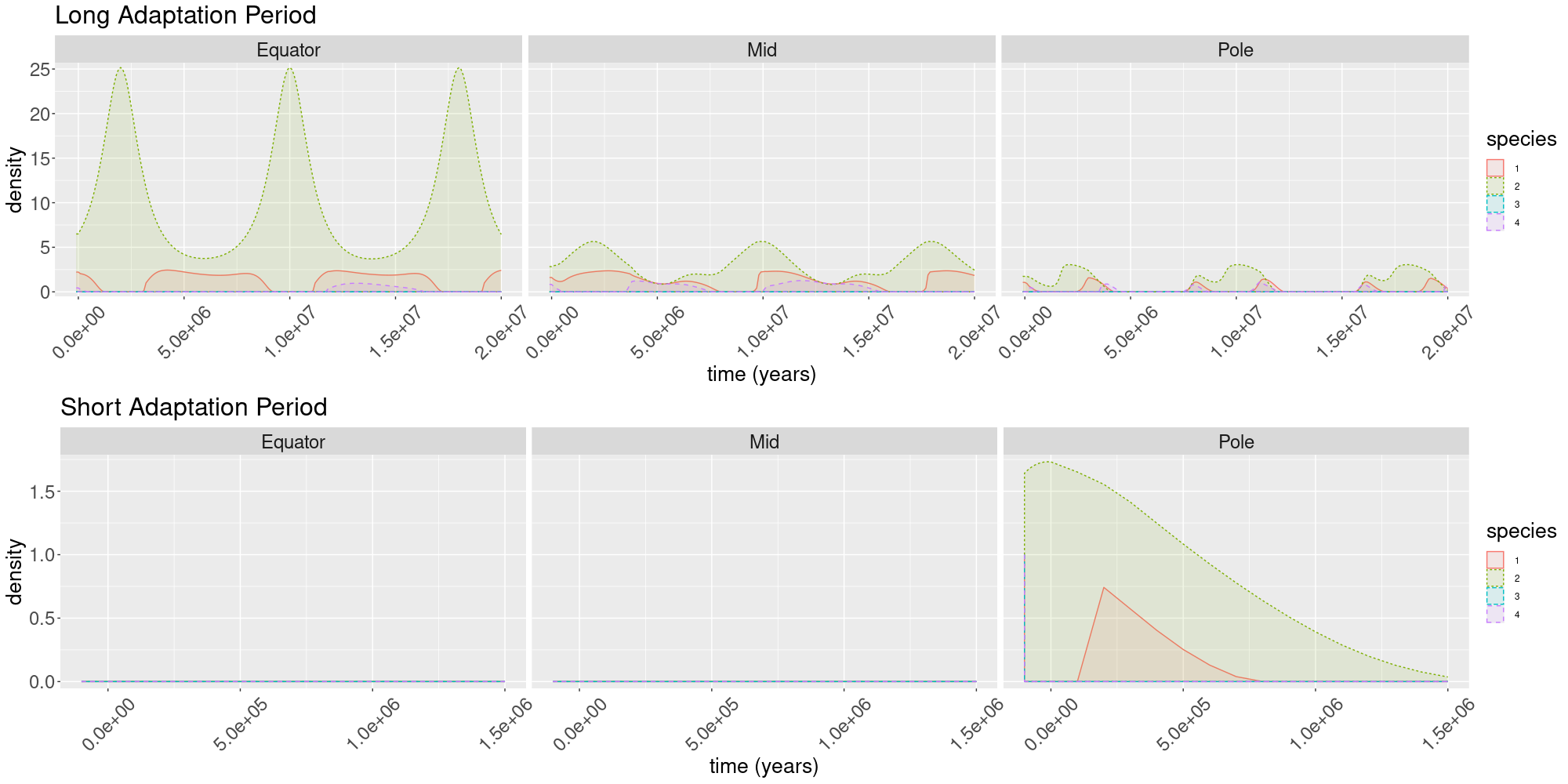}
    \caption{Population density over time in a periodic climate change, for long and short adaptation times}
    \label{fig:periodicContTime}
\end{figure}

Since the spatial distribution in figure \ref{fig:sinDensity} is typically monotonic, the same information can be illustrated in continuous time rather than continuous space, in order to see the periodic behavior more clearly. The alternative display configuration is used in figure \ref{fig:periodicContTime}, when the discrete loci are the same as in figures \ref{fig:stepTemp} and \ref{fig:sinTemp}.

Both simulations are displayed starting from the starting point of the short adaptation case - close to climatic change onset -  since the longer adaptation time population is mostly constant up to that point. 
Notice the panels corresponding to the equatorial and 45$^\circ$ parallel environments are empty on the short adaptation time case since none of the species managed to reach these far habitats before extinction.

\subsection{Secular dynamical evolution induced climate change}
The third case we consider is more realistic, in the sense that we do not follow an arbitrarily chosen non-physical climate change, but consider a potentially realistic model for the dynamical evolution of a planetary system. We consider a two-planet system, with the outer more massive planet orbiting the host star in a highly inclined orbit, leading to the secular evolution of the inner orbit through repeating eccentricity evolution vZLK cycles. In turn, the change in the orbit of the inner terrestrial planet, takes it closer and further away from the host star, thereby changing its insolation by the host star, and hence its surface temperature.     

As mentioned in section \ref{methods}, the hierarchical 3-body system was chosen to have masses that resemble a Sun-Earth-Jupiter system.
The exact parameters that were chosen for the simulation are given in table \ref{tab:sim_params}.

\begin{figure}[ht!]
    \centering
    \includegraphics[width=1\linewidth]{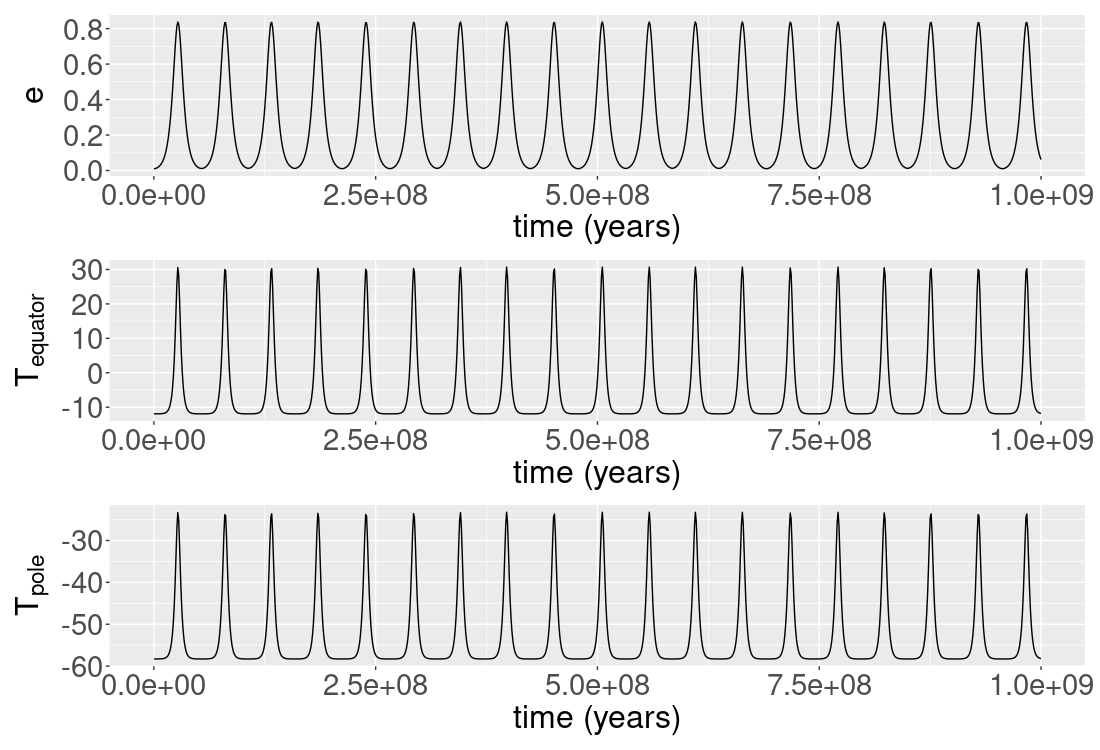}
    \caption{Temperature constraints on the planetary poles and equator over time, for a planet experiencing eccentric vZLK oscillations. The top sub-figure presents the planetary eccentricity over time}
    \label{fig:KozaiTemp}
\end{figure}

\begin{figure}[ht!]
    \centering
    \includegraphics[width=1\linewidth]{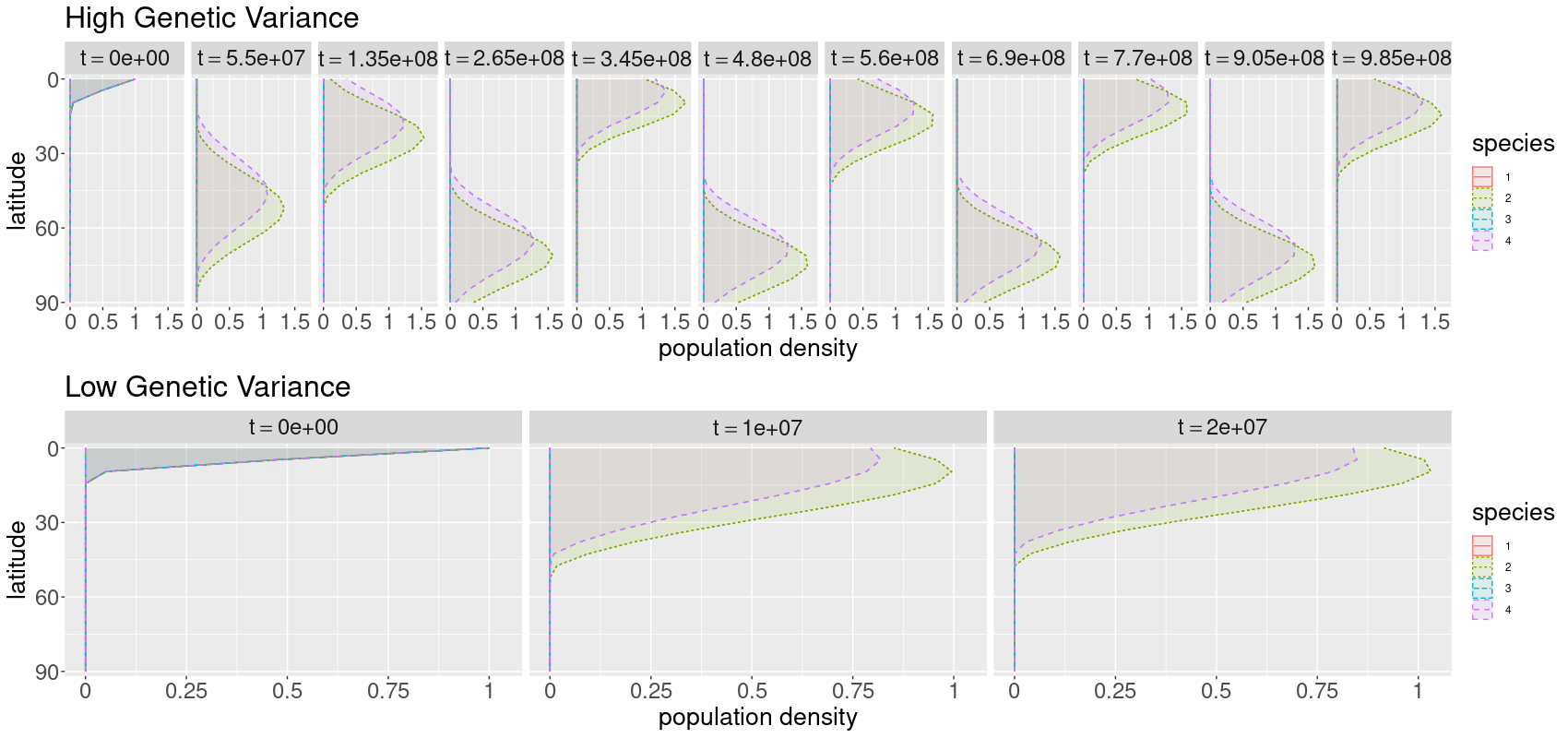}
    \caption{1-D spatial distribution of population density evolving through vZLK-oscillations-dominated climate, for phenotypic variance values above (large) and below (small) the critical variance}
    \label{fig:KozaiDensity}
\end{figure}

\begin{table}[ht]
\centering
\caption{Simulation Parameters}
\label{tab:sim_params}
\begin{tabular}{|l|l|l|}
\hline
\textbf{Parameter} & \textbf{Symbol} & \textbf{Value} \\
\hline
Inner star mass & $m_1$ & $1 \, M_\odot$ \\
Inner planet mass & $m_2$ & $1 \, M_{\earth}$ \\
Outer planet mass& $m_3$ & $7.5 \, M_{Jupiter}$ \\
Inner planet obliquity & $\epsilon$ & $24^\circ$ \\
Inner planet surface albedo& $a$ & $0.29$ \\
Stellar luminosity & $L_s$ & $1 \, L_\odot$ \\
Initial semi-major axis of inner star & $a_1$ & $0.51$ AU \\
Initial semi-major axis of outer star & $a_2$ & $50$ AU \\
\hline
\textbf{Initial vZLK Parameters} & & \\
\hline
Initial eccentricity of inner and outer orbits & $e_1 = e_2$ & $0.01$ \\
Initial inclination of inner orbit & $i_1$ & $64.9^\circ$ \\
Initial inclination of outer orbit & $i_2$ & $0.1^\circ$ \\
Initial arguments of periapsis & $\omega_1 = \omega_2$ & $0$ \\
\hline
\end{tabular}
\end{table}
The system was simulated with no adaptation phase since the vZLK evolution in this case is periodic and generates a substantial climatic shift from the beginning. 
The survival/extinction case separation was expected to result from the relation between the temperature change rate and the species adaptation rate (either dispersal or phenotypic).
For this reason, instead of changing the non-existent preemptive adaptation time, or searching for specific orbital parameters such that extinction and survival behaviors can be demonstrated, we change the the species adaptation parameters. In other words, since the ability to adapt to change effectively depends on the relative timescales for genetic (and/or spatial migration) adaptation and the timescale (and amplitude) of the environmental climate change, the separation between survival and extinction of a species can be demonstrated either by changing the latter, the environmental evolution parameters, as done previously, or as demonstrated here, by changing the former, adaptation capabilities of the species themselves 
For simplicity, the nominal dispersal rate was kept in a relatively high value $d = 1\cdot 10^{-3} (\frac{\pi}{2} \frac{ R_{planet}}{year})$, and only the phenotypic variance was scanned around the approximated critical value. This critical value was derived directly from the maximal temperature change rate.
The resulting temperature profile is depicted in figure \ref{fig:KozaiTemp}, run for 1 billion years. The matching survival/extinction cases around the actual critical phenotypic variance are illustrated in figure \ref{fig:KozaiDensity}. 

The phenotypic variance values used in this case were (nominally over the species) $v = 6.45 \cdot 10^{-5} (C^{\circ})^2$ for the survival case, and $v = 3.23 \cdot 10^{-5} (C^{\circ})^2$ for the extinction case.
As in the previous cases, the temperatures at the ambient habitats are linearly interpolated between the two externally constrained environments.

\section{Discussion} \label{discussion}
Using the full ecological-evolutionary model simulations, we demonstrated the case for a survival-extinction boundary for a single parameter at a time.  
At first, the pre-climatic-change adaptation time was varied. Fairly simplistic temperature profiles were explored: a smooth rising step; and a sinusoidal periodic profile.
As expected, adaptation time does affect the survival capability of species. In our simulations, we addressed evolution shortly after abiogenesis, but the same conclusions apply also to biological communities that hold a rather narrow set of traits at a specific habitat,  that is, not genetically or spatially diverse.
The population density is decreasing constantly through climate change for both cases, until the midpoint.
Around 1 million years into the climate change, $\frac{dT}{dt}$ reaches its maximal value. This is the failure point for the short adaptation case, and the low point for the long adaptation case as well.
Those findings correlate well with the basic model results in figure \ref{fig:firstDemo}. These show that the population density anti-correlates with the temperature rise through the trait lag. Even though all other parameters in the system (i.e relations between $\sigma$,$\rho$,$v$, $\Delta t$ ) were changed and the model had several simplifications relaxed, the dependency between $n$ and $T-m$ is kept. It is straightforward to deduce that the larger $|\frac{dT}{dt}|$ is, compared to the time-constant genetic and dispersal capabilities $v$, $d$, the trait lag will increase, and thereby negative growth will occur, especially when exceeding the temperature tolerance $\sigma$. This emphasizes the role of the phenotypic and dispersal variances as the main coping mechanism with fast climatic variations. 
Further evidence for the expected anti-correlation between population growth and $\Delta T$ can be seen in figures \ref{fig:sinTemp} and \ref{fig:periodicContTime}. All extrema of the population density profile for all depicted locations are found at the times for which $\frac{dT}{dt}$ exhibits extrema.
In the long adaptation time case, the typical timescale for population update is sufficiently short for the population density to tightly follow the climatic trend, without any apparent lags.  
Furthermore, figure \ref{fig:stepDensity} reveals that for the short adaptation time case, the biological communities remained mainly around their starting point at the pole, and eventually failed to adapt to the local climate change. In the other case, where the species were given a long adaptation time before climatic change onset, the species were able to redistribute through the planetary habitats, all the way to the equator, and eventually survive. One should bear in mind that in the simulated case,  $\Delta T$ is unbalanced throughout the planet when the lightest temperature rise $C_{min}$ occurs at the equator, and the harshest one $C_{max}$ at the poles. Over a given time period $\Delta t$, this claim is equivalent to comparing derivatives as before. For this reason, the communities who did not manage to escape the polar harsh-to-be habitat, went extinct, while the equatorial communities exhibited a smaller temperature rise. While the preemptive time is the degree of freedom in this case separation, the survival strategy itself is migration, and eventually a spatial selection occurs which favors the habitats with the more moderate climatic change.
This phenomenon is observable both for the monotonic temperature rise case and the periodic sinusoidal climate change.
However, a more delicate feature is the case shown where already at $t=0$ the equator is populated more than the abiogenesis location at the poles. 
The reason is rooted in our model choice for a trait dependent tolerance. As indicated in section \ref{Full Ecological Model} , warmer habitats get a higher maximal growth rate in exchange to lower tolerance. On simulation with constant tolerance, the early preference for the equator did not appear.
Lastly, the nominal phenotypic variance $v$ was the degree of freedom on a vZLK oscillatory temperature profile.  The population density, regardless of the profile not being completely periodic as before, is still strictly following the temperature trend. In this case, a clear extinction-survival boundary in $v$ was found. In the extinction case - life did not make it through the first period of the climatic change. In the survival case -  a slightly larger $v$ allowed us to establish a better starting point for the population density. From that point onwards, the main coping mechanism is dominantly through migration, since the same density profile appears to shift from one best-suited habitat to another, without allowance for growth under such extreme and rapid climatic change. It should be noticed that for both cases the migration capability $d$ was high and the same, but migration alone does not guarantee survival.
Our models demonstrate how both spatial migration and genetic adaptation coping mechanisms can play a role in survival, where in some cases a single adaptation mechanism is sufficient for survival and in others, both play a key role.
Under harsh and frequent climatic variations as in a high-inclination planet, with a massive perturbing planet nearby, the additional degrees of freedom are contributing to the survival chances of the native bio-communities. 

\section{Summary} \label{summary}
In this work, we aimed to introduce and demonstrate the concept of adaptive habitability and ecological (astroecology) behavior in the context of life on exoplanets. The theoretical model chosen for the task was explored with a highly simplified planet-level behavior, without considering specific details as to what is the ecological or chemical nature of the life in question, nor exploring detailed atmospheric processes that would introduce additional different effects on the climate. A mathematical claim about the capacity of an adapting object to a varying condition will be the more loyal description of this work.
While simplified, we demonstrate how such types of models, to some extent explored in the context of local ecological habitats on Earth, can be adapted and explored in the astro-biological context, and in the frame of secular dynamics evolution of planetary systems. 
We have shown that theoretical life can cope with climate change on other planets through ecologically defined characteristics, with a clear relation between the amount of environmental shift and the adaptation capabilities.
This is a mere proof of concept, which shows habitability can be defined through temporal states, the same as our own planet has changed throughout the evolution of life.
Future directions are plentiful. 
\begin{acknowledgments}
We would like to acknowledge the support from the Minerva center for life under extreme planetary conditions.
\end{acknowledgments}

\section{Code Availability} \label{code}
Computer code for implementing our model and replicating our results can be found at 
https://github.com/itaywe1998/Astro-Ecology-Paper.git
\bibliography{refs.bib}{}
\bibliographystyle{aasjournal}

\end{document}